\documentclass[prd,showpacs,showkeys,preprint]{revtex4}

\usepackage{amsmath}
\usepackage{amssymb}
\usepackage{yfonts}[1988/10/03]
\usepackage{graphicx}
\newcommand {\beq}{\begin{equation}}
\newcommand {\eeq}{\end{equation}}
\newcommand {\beqa}{\begin{eqnarray}}
\newcommand {\eeqa}{\end{eqnarray}}

\begin{document}
\title{Supernova Neutrino Background Bound on the SFR History}
\author{  Jafar Khodagholizadeh $^{1}$\footnote{Lecturer in Shahid Beheshti branch, Farhangian University, Tehran, Iran: gholizadeh@ipm.ir}, Sepehr Arbabi $^{2}$ \footnote{arbabi@ipm.ir} and Zamri Zainal Abidin$^{3}  $\footnote{zzaa@um.edu.my}}
\affiliation{ School of physics, Institute for research in fundamental sciences (IPM), Tehran, Iran.\\
$^{2}$
Qom University of Technology, Physics Department, Qom,  Iran\\ $^{3}$ Department of Physics, Faculty of Science, University of Malaya, 50603 Kuala Lumpur, Malasia.}
\date{\today} 

\begin{abstract}
The purpose of the present study is to compare the predictions of different models of star formation rate (SFR) history  in the universe with the upper limit of Super Kamiokande for the neutrino background. To this aim we have calculated the expected neutrino density for the most popular models of SFR history, Hogg et al., Glazebrook et al., Cole et al., Yuksel et al., Hernquist et al. and Kaplinghat et al. Different from previous studies we have used the $ \Lambda $CDM model with $ \Omega_{\Lambda}=0.7 $. We have assumed that the detector used for the detection the neutrino flux is SuperK and also we have assumed that the electron neutrinos produced in the Supernovae oscillate equally to the three standard model flavors. By these assumptions all models stay below the upper limit of SuperK on the event rate and the detection of the supernova relic neutrino background (SRNB) remains undetected. Future neutrino detectors such as KM3Net will be able to detect the SRNB and distinguish between the models of the SFR history.
\end{abstract}
\maketitle
\section{Introduction}
Neutrino telescopes are right now the only possibility of exploring the distant universe independent of photons. By studying the relic neutrinos stemming from distant supernovae we can derive information about the history of the universe and its star formation rate (SFR). It is a well established fact that core collapse supernovae like Type II SN emit $99\%$ of their energy in the form of neutrinos. These neutrinos, form the so called relic $ \bar{\nu}_{e} $ background, which is created by Type II supernovae and has been detected in large Neutrino detectors such as SuperK, SNOLlab, Icecube and future detectors such as Km3Net. The relic $ \bar{\nu}_{e} $ flux depends on the rate of $ SN_{II}$ rate as function of redshift which itself is a function of the SFR history of the universe. Also the metal enrichment history and the cosmic chemical evolution are closely related to SFR and play an important role in the study of the universe. In the recent past a wealth of methods for estimating the star formation rate history have been applied. For a recent review of the observational methods used in this topic, cf. Madau, Wilkinson 2014 \cite{Madau14}. There is a general consensus that the star formation increases from now to redshift unity and decreases exponentially towards higher redshifts and earlier times.

The star formation history of the universe was the subject of a pioneering work by Madau et al.96 \cite{Madau96}. Sicne then a large number of studies have been undertaken until the recent past. For example Pei and Fall compared models of cosmic chemical evolution with observational indicators, the abundance of neutral hydrogen, heavy elements and the dust on damped $Ly\alpha$ systems and present-day galaxies \cite{Pei,Fall}. Totani et. al \cite{Totani, Totani1} used a time-dependent supernova rate from a model of galaxy evolution based on the population synthesis method. In a similar way, in the work of Bisnovatyi-Kogan and Seidev \cite{Bisnovatyi}, galaxy evolution is considered and it is assumed that the supernova rate depends redshift proportional to $(1+z)^{A}$. In their model, the supernova rate is much higher in the early phase of elliptical galaxies, and more than half of total supernova explode during the initial 1 $Gyr $ after the formation of galaxies. Thereafter the supernovae rate in spiral galaxies becomes dominant and the total number of supernova until the present is consistent with the requirements of nucleosynthesis. Also Hartman and Woosley \cite{Hartmann} compared a model of cosmic chemical evolution with observations of $Ly\alpha$ systems and faint galaxy surveys. They obtain a a power-law SFR with a redshift dependence (SFR $\propto t^{-2.5}$\cite{Lilly}). In most models the maximum of SFR takes place at a redshift of order unity, where overall rate was about 10 times higher than it is today.

In another analysis of the SFR, Hogg considered measurements of radio, infrared and ultraviolet broad-band photometric indicators, visible and near-ultraviolet line-emission indicators from redshift unity to the present day \cite{Hogg}. Assuming that SFR is proportional to $(1+z)^{\beta}$, the best-fit exponent was obtained $\beta=2.7\pm0.7$
Cole et. al  measured galaxy luminosity functions in the near infrared from a combined 2MASS-2dFGRS selected galaxy catalogue \cite{Cole}. On this basis they determined the mass of stars formed until today and obtained $\dot{\rho}_{\star}=(a+bz)/[1+(z/c)^{d}]h M_{\bigodot}yr^{-1} Mpc^{-3}$ where$(a,b,c,d)=(0.0166,0.1848,1.9474,2.6316)$.
Hernquest and Springel use analytical physical reasons to model SFR.,$\dot{\rho}_{\star}$ \cite{Hern}. They found that at early times $\dot{\rho}_{\star}$ generically rises exponentially as $z$ decreases, independent of the details of the physical model for star formation, but dependent on the normalization and shape of the cosmological power spectrum. They conclude that at lower redshifts, the star formation rate scales approximately as $\dot{\rho}_{\star}\propto H(z)^{4/3}$. In this model the peak of SFR depends on the model parameters but half of the stars have formed at redshifts higher than $z\simeq2.2$.
Glazebrook and et.al study the overall spectrum of galaxies obtained from the red-selected Sloan Digital Sky Survey and compared the results with the blue-selected 2dF Galaxy redshift Survay \cite{Glaze}. Here they used a double power-law parametrization of the SFR  with a break at redshift unity: SFR $\propto (1+z)^{\beta}$ for $z<1$ and $\propto (1+z)^{\alpha}$ for $1<z<5$  and star formation rate is zero  for $z>5$..
Finally Yuksel and et al studied SFR in relation to the Gamma-ray bursts \cite{Yuksel}. The Gamma-ray bursts have the advantage that they can be observed at higher redshifts and determine the SFR at $z=4-7$. The result for SFR reported there was that a steep drop exist in the SFR up to at least $z\sim 6.5$

In the present study we have chosen analytic models mentioned above for studying the SFR. Here we start out with a model which has a few parameters which can be constrained in comparison of such models with observations, which is the relic supernova background in our case. This paper is organized as follows. In the next section we calculate the supernova relic neutrino density for the different models of SFR. In section 3, we determine the predictions of the models for the neutrino event rate at the SuperKamiokande detector. The results are presented in section 4, followed by a short discussion in section5.

\section{The Supernovae Relic Neutrino Spectrum}\label{S2}
If the supernova rate per unit commoving volume at redshift $ z$ is $ N_{SN}(Z) $ and the neutrino energy distribution at the source (at energy $ \epsilon $) is $ L_{\nu}^{S}(\epsilon) $, then the expected flux of relic neutrinos on Earth is given by \cite{Kaplin}:
\begin{eqnarray}\label{1}
j_{\nu}(\epsilon)=\frac{C}{H_{0}}\int dz\frac{N_{SN}(z)<L^{S}_{\nu} (\epsilon')>}{(1+z)\sqrt{\Omega_{\Lambda}+(1+z)^{3}\Omega_{M}}}
\end{eqnarray}

where $ \epsilon'= (1+z)\epsilon $ is the rest frame neutrino energy and the neutrinos are assumed to be massless.
The spectrum of the neutrinos is parameterized as a Fermi-Dirac distribution with zero chemical potential and normalized to the total energy in a particular neutrino species $ (E_{\nu}) $ emitted by the supernova, i.e. $  \int L_{\nu}^{S}(\epsilon) \epsilon d\epsilon=E_{\epsilon}$. For each neutrino species $\nu_i$ the energy distribution is given by:
\begin{equation}
L^{S}_{\nu_i}=E_{\nu_i} \times \frac{120}{7\pi^{4}}\frac{\epsilon'^{2}}{T_{\nu_i}^{4}}[\exp(\frac{\epsilon'}{T_{\nu_i}})+1]^{-1}
\end{equation}
From the determination of the SFR we extract the SN rate:
\begin{equation}
N_{SN}(z)\propto \dot{\rho}(z)
\end{equation}
where $ \dot{\rho}(z) $ is the star formation rate. Averaging over a Salpeter Initial Mass Function (IMF) for $ M >8M_{\odot} $, the supernova rate is $ N_{SN}(z)=(\frac{0.013}{M_{\odot}})\dot{\rho}(z) $Ì‡, while the star formation rate is measured in solar masses.
Following  \cite{Totani,Totani1,Bisnovatyi,Hartmann,Hogg,Baldry}, we parameterize the SFR as:

\begin{equation}
\dot{\rho}(z)\propto (1+z)^{A}
\end{equation}
where the constant $A$ has a different value for $ z<1 $ and $ 1<z<2 $. We also assume that the behavior of $A$ at $ 1<z<2 $, continues to higher redshifts. For $z<1$ Hogg \cite{Hogg} has compiled measurements of the UV and $H_\alpha $ luminosity density and obtained the $ 68 \% $ C.L limits of $ A=2.7\pm0.7 $ for $\Lambda CDM$ model and $ A=3.3\pm0.8 $ for $\Omega_{M}=1$ model. From the optical spectrographic measurements in the Sloan Digital Sky Survey (SDSS) limits on A were found to be $ 2-3 $ for $ z<1 $ and $ 0-1 $ for $ z>1 $, \cite{Glaze}. With these assumptions the rate of relic neutrinos on earth will be:
\begin{eqnarray}
J_{\nu}=\frac{C}{H_{0}} \frac{120}{7\pi^{4}}\frac{<E_{\nu}>}{<T_{\nu}>^{4}}\frac{0.013}{M_{\odot}}  \int dz\frac{(1+z)^{A} \epsilon '^{2}}{(1+z)\sqrt{\Omega_{\Lambda}+ (1+z)^{3}\Omega_{M}}}\frac{1}{\exp(\frac{\epsilon'}{T_{\nu}})+1}
\end{eqnarray}
As before $ \epsilon'=(1+z)\epsilon $. By setting $ x=1+z $ , we can rewrite the integral:
\begin{equation}
\label{nuflux}
J_{\nu}=\frac{C}{H_{0}} \frac{120}{7\pi^{4}}\frac{<E_{\nu}>}{<T_{\nu}>^{4}}\frac{0.013}{M_{\odot}} \epsilon ^{2}\int_{1}^{\infty} dx\frac{x^{A+1} }{\sqrt{\Omega_{\Lambda}+ x^{3}\Omega_{M}}}\frac{1}{\exp(\frac{\epsilon}{T_{\nu}})+1}
\end{equation}
The data from $ SN1987 A $ gave $ E_{\nu}= 8 \times 10^{52} ergs $  and $ T_{\nu}=4.8  $Mev, while the model introduced by Woesley et al\cite{woos} predicts that for a $ 25M\odot $ supernova progenitor: $ E_{\nu}= 11\times 10^{52}erg $, $ T_{\nu}=5.3  $Mev.
As an approximate solution for (\ref{nuflux}) we find:
\begin{eqnarray}
J_{\nu}=\frac{C}{H_{0}} \frac{120}{7\pi^{4}}\frac{<E_{\nu}>}{<T_{\nu}>^{3}}\frac{0.013}{M_{\odot}} \frac{\epsilon}{\Omega}_{M} [1.8\sqrt{\Omega_{\Lambda}+8\Omega_{M}}-0.5\sqrt{\Omega_{\Lambda}+9\Omega_{M}}]
\end{eqnarray}
 But the SFR is not always simple and it has a more complex shape in the general case. Cole et.al obtained from high-z galaxies and gamma-ray bursts data,assuming a Salpeter initial mass function  \cite{Cole}
\begin{eqnarray}
\dot{\rho}(z)=(a+bz)h/[1+(z/c)^{d}]
\end{eqnarray}

with $a=0.0389 $ ,$b=0.0545 $ , $c=2.973 $ and $d=3.655 $ . Hermquist and Springel found \cite{Hern}
\begin{eqnarray}
\dot{\rho}(z)=\dot{\rho}_{0}\chi^{2}/[1+\alpha(\chi -1)^{3}\exp(\beta \chi^{7/4})]
\end{eqnarray}
where $ \chi=[H(z)/H_{0}]^{2/3} $ , $ \dot{\rho}_{0}=0.030 $ ,$ \alpha=0.323 $ and $ \beta=0.051 $. In the Yuksel et.al model, the SFR is \cite{Yuksel}
\begin{eqnarray}
\dot{\rho}(z)=\dot{\rho}_{0}[(1+z)^{\eta\eta}+\lbrace(1+z)/B\rbrace^{\beta\eta}+\lbrace(1+z)/C\rbrace^{\gamma\eta}]^{1/\eta}/[1+\alpha(\chi -1)^{3}\exp(\beta \chi^{7/4})]
\end{eqnarray}
where $ B=(1+z_{1})^{1-\alpha/\beta} $, $ C=(1+z_{1})^{(\beta-\alpha)/\gamma}(1+z_{2})^{1-\beta/\alpha} $ while $ \dot{\rho}_{0}=0.0285 $, $ \alpha=1.6 $,  $ \beta=-1.2 $, $ \gamma=-5.7 $ in which $ z_{1}=1.7 $, $ z_{2}=5.0 $ and $ \eta=-1.62 $. Therefore the relation (\ref{1}) becomes
\begin{equation}\label{2}
J_{\nu}=\frac{C}{H_{0}} \frac{120}{7\pi^{4}}\frac{<E_{\nu}>}{<T_{\nu}>^{4}}\frac{0.013}{M_{\odot}} \epsilon ^{2}\int_{1}^{\infty} dx\frac{x \dot{\rho}(x) }{\sqrt{\Omega_{\Lambda}+ x^{3}\Omega_{M}}}\frac{1}{\exp(\frac{\epsilon}{T_{\nu}})+1}
\end{equation}
Neutrino density function for each model has been plotted in figure1. The results presented here in section .4 are based on numerical calculations of (\ref{nuflux}) and (\ref{2}) in the intervals $z=0$ to $ z=\infty$ or $x=1$ to $ x=\infty$.
\begin{figure}
\centering
\includegraphics[scale=0.5]{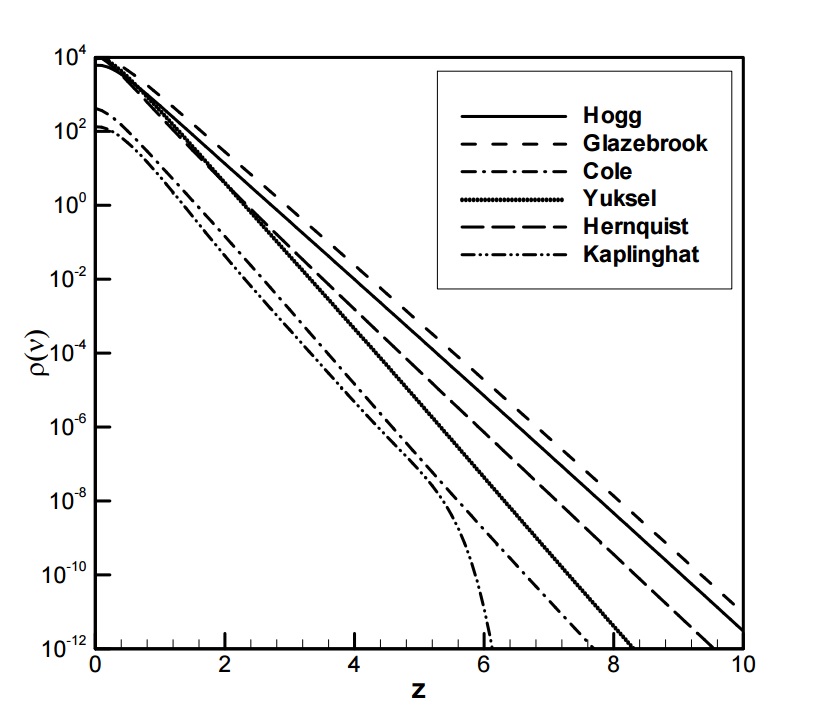}
\caption{Neutrino density predictions at different redshifts for each model as explained section 2.}
\label{fig:3}
\end{figure}
\section{Event Rate at Super Kamiokande Detector}
State of the art neutrino telescopes are water Cerenkov detectors, where two general types can
be distinguished. The dominant reaction in a light water Cerenkov detector such as SuperK is $ \bar{\nu}_{e}p\longrightarrow ne^{+} $ with the cross section $ \delta_{\epsilon}(e) $ two order of magnitude larger than that of the scattering reaction $ \nu_{e}e\longrightarrow\nu_{e}e $  in a heavy water Cerenkov detector such as SNO is $ \bar{\nu}D\longrightarrow nn e^{+} $, where the cross section of these reactions are denoted $\sigma_{i} $. $ i $ is a positron at SuperK type of detector and Deuterium at the SNO type. For simplicity at the calculation of the event rate $R$, the efficiency of the detectors are assumed to be $ 100\% $ in the observable energy window. SuperK does not detect supernovae relic neutrinos at all energies. Below $ 10 $ Mev, the $ \bar{\nu}_{e} $ flux is dominated by nuclear reactors and other sources of $ \bar{\nu}_{e} $ arriving on Earth  are not distinguishable. Between  $ 10 $ Mev and $ 19 $ Mev the Neutrino background is due to solar neutrinos and due to Cosmic muons in detectors. Between $ 19 $ Mev and nearly $ 20.3 $Mev the background is due to atmospheric neutrinos and the flux of neutrinos at energies greater than $ 36 $ Mev rapidly falls exponentially, so that the observable flux for neutrinos in SuperK detector is from $20.3$ to $36.3 $ Mev where $ \epsilon=E_{e}+1.3 Mev $ and $ E_{e} $ is the energy of positron and $ 1.3 $ Mev the neutron-positron mass difference.
Therefore the differential event rate in the internal $d\epsilon $ at SuperK is
\begin{eqnarray}
R_{superK} = B(1.51\times 10^{33})(9.52 \times 10^{52})\frac{<E_{\nu}>}{<T_{\nu}>^{4}}(\frac{0.013}{M_{\odot}}) \int_{0}^{\infty} \frac{\dot{\rho}(z)}{(1+z)\sqrt{\Omega_{M}(1+z)^{3}+\Omega_{\Lambda}}}~~~~~~~~ \nonumber\\
 \times \int_{20.3}^{36.3} \frac{\epsilon ^{2}(\epsilon-1.3)^{2}d\epsilon} {\exp(\frac{\epsilon x}{<T_{\nu}>})+1}\nonumber\\ =
B(1.51\times 10^{33})(9.52 \times 10^{52})\frac{<E_{\nu}>}{<T_{\nu}>^{4}}(\frac{0.013}{M_{\odot}}) \int_{1}^{\infty} \frac{x \dot{\rho}(x)dx}{\sqrt{\Omega_{M}x^{3}+\Omega_{\Lambda}}}  \times \int_{20.3}^{36.3} \frac{\epsilon ^{2}(\epsilon-1.3)^{2}d\epsilon} {\exp(\frac{\epsilon x}{<T_{\nu}>})+1}\nonumber\\
\end{eqnarray}

where $ B=(\frac{120}{7\pi^{4}})C H_{0} ^{-1}=1056 h_{50}^{-1}$ , we use $ \sigma_{p}(\epsilon)= 9.52 \times 10^{52} E_{e}P_{e }cm^{2}$   \cite{vogel} ,$ <E_{\nu}>=11 \times 10^{52} $ergs and $<T_{\nu}>=5.3$Mev. So the differential event rate in the interval $ d\epsilon $ is $ N_{p}\sigma_{p}J_{\nu}(\epsilon) d\epsilon $ and the predicted event rate at the detector are given by:
\begin{equation}
R=B N_{i}\frac{<E_{\nu}>}{<T_{\nu}>^{4}}(\frac{0.013}{M_{\odot}})\int \frac{x \dot{\rho}(x)~dx }{\sqrt{\Omega_{\Lambda}+\Omega_{M}x^{3}}}\int \frac{\epsilon^{2}\sigma_{i}(\epsilon)}{\exp(\frac{x\epsilon}{<T_{\nu}>})+1}
\end{equation}
The $d\epsilon$ delineate the energy window between 20.3 to 36.3 and $ N_{p} $ is the number of free protons in SuperK detector with a sensitive water mass of $ 22.5 $ Ktons and $ N_{p}=1.51 \times 10^{33} $ . The SN relic $ \bar{\nu}_{e} $ event rate at SuperK in $ \Lambda CDM $ model can be written as
\begin{eqnarray}
R=0.063(\frac{M_{\odot}}{\langle M_{z}\rangle})(\frac{\langle E_{\nu}\rangle}{10^{53} ergs})(\frac{\langle T_{\nu}\rangle}{Mev})\frac{events}{22.5 kton-yaer}
\end{eqnarray}
We have set $ h_{50}=1 $ and also the average metal yield per supernova have taken to be $ 1 M_{\odot} $ in the interest of obtaining  an upper bound to the event rate, while the first number in the right side of the above expression with $ \Omega_{M}=1 $ is  equal to 0.066.

More recent estimates by Totani et al \cite{Totani}, using the population synthesis method to model the evolution of star formation in galaxies, obtained a prediction for the flux of SRN at superK (in the energy interval from 15 to 40 Mev) of $ 1.2 yr^{-1} $ and the "most optimistic" prediction for their model was an event rate of $ 4.7 yr^{-1} $. Malaney \cite{Malaney} used the Pei and Fall result \cite{Pei} in order to parametrize the evolution of the Cosmic gas density rate and integrated over all energies and found a total SRN flux of $ 2.0-5.4 cm^{-2}s^{-1} $. Hartmann and Woosley \cite{Hartmann} used an SN rate proportional to $ (1+z)^{4} $ and their best estimation was $ \sim 0.2 cm^{-2} sec^{-1} $ . Kaplinghat et al. \cite{Kaplin} used the assumption that the supernova rate tracks the metal enrichment rate  can be written as: $ N_{SN}(z)=\frac{\dot{\rho _{z}}(z)}{<M_{z}>} $ where $ <M_{z}> $ is the average yield of " metal" $ ( z > 6 )$  per supernova and $ \dot{\rho}_{z} $ is the metal enrichment rate per unit comoving volume $ 22.5  Kton - year $ and the neutrino flux at superK is $ 1.6 cm^{-2}sec^{-1} $ or event rate is $ R<4  $events for $ 19 MeV< E_{e}<35 MeV$ and over all energies the event rate at SuperK is $ 54 ~cm^{-2} sec^{-1} $.

\section{Results }
Here to obtain the most optimistic of SRN event rate at SuperK we consider neutrino oscillation as  a mechanism for maximizing the SRN flux. We have assumed that $ \Omega_{M}=0.3 $ and $ \Omega_{\Lambda}=0.7 $. In Kaplinghat et al model with  $ \Omega_{\Lambda} $, our estimates gives the neutrino flux at superK is $ 1.54 cm^{-2}sec^{-1} $and the event rate is $ R<3 $ events. For Hogg's model wich has compiled the UV and $H_{\alpha}$ luminosity density and obtained the SFR as $(1+z)^{2.7\pm 0.7}$ the flux at detector is 108.57 $cm^{-2}s^{-1}$. For the SFR obtained from the optical  spectrographic measurements in the Sloan Digital Sky Survey (SDSS) the maximum flux of SRN is 111.43 $cm^{-2}s^{-1}$ and minimum flux is 91.1 $cm^{-2}s^{-1}$. With the Cole model the flux is 3.02 $cm^{-2}s^{-1}$ and Yuksel model the flux is 80.32 $cm^{-2}s^{-1}$ and then for Hernquest model the flux at detector will be 70.13 $cm^{-2}s^{-1}$. So then their upper bound from models of Fall, Hogg, Glazebrook, Cole, Yuksel and Hernquest are 3 event, 212 event, 217 event, 6 event, 156 event and 136 event respectively and as summarized in the table \ref{tab}.
\begin{table}
\begin{center}
\begin{tabular}{c | c | c }\hline
Star Formation Rate  ~~&      ~~The neutrino flux ($cm^{-2} sec^{-1}$)~~ & ~~Event Rate( $R <$ )\\ \hline
Kaplinghat's model &  1.54& 3\\
Hogg 's model& 108.57 &212\\
Glazebrook's model &  111.43&217\\
Cole's model&  3.02&6\\
Yuksel's model&  80.32&156\\
Hennguiest's model&  70.13&136\\
\end{tabular}
\end{center}
\caption{The neutrino flux  and upper bound for SRN at superK in $\Lambda CDM $ cosmology\label{tab}}
\end{table}
\section{Discussion}
In the present work we have studied the predictions of the different models of SFR for the number of expected relic neutrinos in large neutrino detectors such as SuperK. These relic neutrinos have been produced at every supernova explosion throughout the history of the universe, which transfer almost all their energy to neutrinos. Here we have calculated the neutrino production for different SFR models and have found that the models of Glazebrook \cite{Glaze} and Hogg \cite{Hogg} yield an estimate for the number of  relic neutrinos closer to the lower limit presently given by SuperK.
As the relic neutrino background is an independent way of constraining the SFR of the universe, this method can be regarded as complimentary to the more standard methods. As the construction of more modern neutrino telescopes is underway, the predictions of the SFR models for the relic neutrinos can be soon tested with much more and better data of these telescopes. Regarding this perspective the present study can be expected to be extended soon and yield more fruitful results in the light of new data.
\section*{Acknowledgement}
The author Abidin Z.Z. would like to acknowledge the University of Malaya's  HIR grant UM.S/625/3/HIR/28 for their funding.

\end{document}